# Genome-wide scan of 29,141 African Americans finds no evidence of selection since admixture

Gaurav Bhatia[1,2], Arti Tandon[2,3], Melinda C. Aldrich[4,5], Christine B. Ambrosone[6], Christopher Amos[7], Elisa V. Bandera[8], Sonja I. Berndt[9], Leslie Bernstein[10], William J. Blot[4,11], Cathryn H. Bock[12], Neil Caporaso[9], Graham Casey[13], Sandra L. Deming[4], W. Ryan Diver[14], Susan M. Gapstur[14], Elizabeth M. Gillanders[15], Curtis C. Harris[16], Brian E. Henderson[13], Sue A. Ingles[13], William Isaacs[17], Esther M. John[18], Rick A. Kittles[19], Emma Larkin[20], Lorna H. McNeill[21], Robert C. Millikan[22, †], Adam Murphy[23], Christine Neslund-Dudas[24], Sarah Nyante[22], Michael F. Press[13], Jorge L. Rodriguez-Gil[25], Benjamin A. Rybicki[24], Ann G. Schwartz[12], Lisa B. Signorello[4,11], Margaret Spitz[7], Sara S. Strom[26], Margaret A. Tucker[9], John K. Wiencke[27], John S. Witte[28], Xifeng Wu[7], Yuko Yamamura[26], Krista A. Zanetti[16,15], Wei Zheng[4], Regina G. Ziegler[9], Stephen J. Chanock[9], Christopher A. Haiman[13], David Reich[2,3, *] and Alkes L.. Price[2,29,*]

[1]Harvard–Massachusetts Institute of Technology (MIT), Division of Health, Science, and Technology, Cambridge, Massachusetts 02139, USA
[2]Broad Institute of MIT and Harvard, 7 Cambridge Center, Cambridge, MA 02142, USA
[3]Dept. of Genetics, Harvard Medical School, New Research Bldg., 77 Ave. Louis Pasteur, Boston, MA 02115, USA
[4]Division of Epidemiology in the Department of Medicine, Vanderbilt Epidemiology Center; and the Vanderbilt-Ingram Cancer Center, Vanderbilt University School of Medicine, Nashville, TN 37203, USA
[5]Department of Thoracic Surgery, Vanderbilt University School of Medicine, Nashville, Tennessee 37203, USA
[6]Department of Cancer Prevention and Control, Roswell Park Cancer Institute, Buffalo, NY 14263, USA
[7]Department of Epidemiology, Division of Cancer Prevention and Population Sciences, The University of Texas MD Anderson Cancer Center, Houston, Texas 77030, USA
[8]The Cancer Institute of New Jersey, New Brunswick, NJ 08903, USA
[9]Division of Cancer Epidemiology and Genetics, National Cancer Institute, Bethesda, MD 20892, USA
[10]Division of Cancer Etiology, Department of Population Sciences, Beckman Research Institute, City of Hope, CA 91010, USA
[11]International Epidemiology Institute, Rockville, MD 20850, USA
[12]Karmanos Cancer Institute and Department of Oncology, Wayne State University of Medicine, Detroit, Michigan 48201, USA.
[13]Department of Preventive Medicine and Department of Pathology, Keck School of Medicine, University of Southern California/ Norris Comprehensive Cancer Center, Los Angeles, CA 90033, USA
[14]Epidemiology Research Program, American Cancer Society, Atlanta, GA 30303, USA
[15]Division of Cancer Control and Population Sciences, National Cancer Institute, Bethesda, Maryland 20892, USA
[16]Laboratory of Human Carcinogenesis, Center for Cancer Research, National Cancer Institute, Bethesda, Maryland 20892, USA
[17]James Buchanan Brady Urological Institute, Johns Hopkins Hospital and Medical Institutions, Baltimore, MD 21287, USA
[18]Cancer Prevention Institute of California, Fremont, CA 94538; and Stanford University School of Medicine and Stanford Cancer Center, Stanford, CA 94305, USA
[19]Department of Medicine, University of Illinois at Chicago, Chicago, IL 60607, USA
[20]Department of Medicine, Division of Allergy, Pulmonary and Critical Care, 6100 Medical Center East, Vanderbilt University Medical Center, Nashville, TN 37232-8300, USA
[21]Department of Health Disparities Research, Division of OVP, Cancer Prevention and Population Sciences, and Center for Community Implementation and Dissemination Research, Duncan Family Institute, The University of Texas MD Anderson Cancer Center, Houston, Texas 77030, USA
[22]Department of Epidemiology, Gillings School of Global Public Health, and Lineberger Comprehensive Cancer Center, University of North Carolina, Chapel Hill, NC 27599, USA

[23]Department of Urology, Northwestern University, Chicago, IL 60611, USA
[24]Department of Public Health Sciences, Henry Ford Hospital, Detroit, MI USA
[25]Sylvester Comprehensive Cancer Center and Department of Epidemiology and Public Health, University of Miami Miller School of Medicine, Miami, FL 33136, USA
[26]Department of Epidemiology, The University of Texas M. D. Anderson Cancer Center, Houston, TX 77030, USA
[27]University of California San Francisco, San Francisco, California 94158, USA
[28]Institute for Human Genetics, Departments of Epidemiology and Biostatistics and Urology, University of California, San Francisco, San Francisco, CA 94158, USA
[29]Departments of Epidemiology and Biostatistics, Harvard School of Public Health, Boston, MA 02115, USA
† In memoriam
* Co-senior authors.

Correspondence to G.B. (gbhatia@mit.edu) or D.R. (reich@genetics.med.harvard.edu)

## Abstract

We scanned through the genomes of 29,141 African Americans, searching for loci where the average proportion of African ancestry deviates significantly from the genome-wide average. We failed to find any genome-wide significant deviations, and conclude that any selection in African Americans since admixture is sufficiently weak that it falls below the threshold of our power to detect it using a large sample size. These results stand in contrast to the findings of a recent study of selection in African Americans. That study, which had 15 times fewer samples, reported six loci with significant deviations. We show that the discrepancy is likely due to insufficient correction for multiple hypothesis testing in the previous study. The same study reported 14 loci that showed greater population differentiation between African Americans and Nigerian Yoruba than would be expected in the absence of natural selection. Four such loci were previously shown to be genome-wide significant and likely to be affected by selection, but we show that most of the 10 additional loci are likely to be false positives. Additionally, the most parsimonious explanation for the loci that have significant evidence of unusual differentiation in frequency between Nigerians and Africans Americans is selection in Africa prior to their forced migration to the Americas.

## Introduction

Admixed populations such as African Americans and Latinos are formed by the mixing of populations from different continents. Alleles that are highly differentiated between the ancestral populations and advantageous in the admixed population are expected to rise in frequency after admixture, causing a deviation in local ancestry compared with the genome-wide average (Seldin et al. 2011). This signal can be used to test for selection since admixture.

A recent study applied this approach to 1,890 African Americans (Jin et al. 2012). The study reported six loci as likely targets of natural selection since admixture. However, that study used a genome-wide significance threshold of $P < 2.7 \times 10^{-3}$, correcting for ~20 hypotheses tested. Based on the scale of admixture linkage disequilibrium in African Americans, a more appropriate threshold would be $P < 10^{-5}$, correcting for 5,000 hypotheses tested as recommended by Seldin et al. (2011).

To revisit the issue of whether there is evidence of natural selection since admixture in African Americans, we scanned through the genomes of 29,141 African Americans, using exactly the same genotyping data set that had previously been used to study the landscape of recombination (meiotic crossover) in African Americans (Hinch et al. 2011). This is the largest sample size analyzed to date for this type of study. Using a genome-wide significance threshold of $P < 10^{-5}$, we find no genome-wide significant signals of selection since admixture. The six previously reported loci do not attain nominal significance (P < 0.05), suggesting that they are false positives due to insufficient correction for multiple tests in the previous study.

We also evaluated the 14 signals of unusual population differentiation between African Americans and Yoruba reported by Jin et al. (2012). Four of these loci were previously shown to be genome-wide significant (Bhatia et al. 2011; Ayodo et al. 2007). However, we show that most of the 10 remaining loci are likely to be false positives due to biases that arise when using the Weir and Cockerham (1984) estimator of $F_{ST}$ to compare two populations of very unequal sample size, or due to an insufficient correction for multiple testing. Additionally, at loci with robust signals of selection, the selection is most likely to have occurred within Africa, prior to the arrival of Africans in the Americas. Thus, any conclusions of selection since the arrival of Africans in the Americas should be viewed with caution, and indeed, at present no unambiguous examples of such selection have been empirically documented.

## Results

*Genome-wide scan of 29,141 African Americans*

We performed an admixture scan for unusual deviations in local ancestry in 29,141 African Americans from five cohorts, genotyped on three different platforms (see Methods). We used the HAPMIX software (Price et al. 2009b) to infer local ancestry in each individual and averaged local ancestry estimates across individuals (see Methods). To search for signals of selection since admixture, we computed the difference between the average local ancestry estimate at each locus and the genome-wide average, divided by the empirical standard deviation in local ancestry estimates across SNPs. It is important to divide by the empirical standard deviation, rather than by the theoretical standard deviation expected if all individuals are independent, as in practice there may be cryptic relatedness among samples—as well as systematic error in the ancestry inference—that will inflate the variance across loci compared with what is theoretically expected (see Methods).

The average genome-wide estimate of European ancestry over all samples in the dataset is 0.204 with a standard deviation of 0.117 across individuals and 0.0036 across SNPs. We considered any deviation in local ancestry greater than 4.42 s.d. (i.e. greater than 0.0154) to be genome-wide significant, corresponding to a threshold of $P < 10^{-5}$ (correcting for 5,000 hypotheses tested) as recommended by Seldin et al. (2011). No locus achieved this threshold of genome-wide significance (see Figure 1).

A previous study of selection in 1,890 African Americans (Jin et al. 2012) reported six loci that passed a (less stringent) significance threshold of $P < 2.7 \times 10^{-3}$ (equivalent to a Bonferroni correction for ~20 hypotheses tested). The six loci did not replicate at nominal significance (P < 0.05) in our much larger dataset (see Table 1 and Figure 1), and are likely to be false positives due to an insufficient correction for multiple tests in the previous study. For 5 of the 6 loci in Table 1, the deviation that we observe has the same sign as the deviation reported by Jin et al. This could be due either to statistical chance (P=0.11; 1-sided Fisher's exact test) or small systematic deviations in local ancestry inference that are correlated between the two analyses (see Supplementary Note). In either case, our results show that the proportion of African ancestry at these six loci is not likely to have been strongly affected by natural selection since admixture.

*Inferring selection using allele frequency differences*

Studies of selection often rank single SNP estimates of $F_{ST}$ and report the most highly differentiated SNPs as signals of selection (Akey 2009; McEvoy et al. 2009; Pickrell et al. 2009; Teo et al. 2009; Jin et al. 2012; Akey et al. 2002). These estimates are most often produced using the Weir and Cockerham (1984) (WC) $F_{ST}$ estimator (see Methods). However, a concern with the use of the WC estimator for this application is that estimates can depend on the sample sizes used, potentially resulting in overestimates of the degree of differentiation at single SNPs (Bhatia et al. 2013). In the situation most prone to overestimation, which would be a study of rare variants with large differences in sample sizes between populations, greater than 99% of the highest single SNP $F_{ST}$ estimates would be expected to be the result of inflation due to unequal sample sizes (Figure S1 of Bhatia et al. (2013)). On the other hand, the Hudson estimator (Hudson et al. 1992; Bhatia et al. 2013), which is a simple average of the population-specific estimators of Weir and Hill (2002), does not have this bias.

We tested the magnitude of inflation of WC estimates in real data by reanalyzing the most highly differentiated SNPs reported in a recent analysis of this type (Jin et al. 2012) (see Table 2). This study compared African segments of 1,890 African Americans (AAF) and 113 Yoruba (YRI) at SNPs with MAF > 5%, and reported 40 SNPs—the 99.99[th] percentile of 401,559 SNPs tested—that have $F_{ST}$ greater than 0.0452. These 40 SNPs are clustered into 14 regions, of which 10 are previously unreported targets of natural selection and 4 were reported as genome-wide significant in the parallel study of (Bhatia et al. 2011) (or as nearly genome-wide significant in the case of HBB, a previously identified target of selection (Ayodo et al. 2007)). Of the 10 novel signals, 9 produce lower estimates when we used the Hudson estimator and 3 fall below the Jin et al. (2012) threshold ($F_{ST} > 0.0452$) (see Table 2). We note that the 99.99[th] percentile of $F_{ST}$ could change when switching from the WC estimator to the Hudson estimator. In our analysis, the magnitude of this change was smaller than the decreases observed (see Methods), suggesting that inflated WC $F_{ST}$ estimates may lead to false positive signals of selection.

In addition to issues of $F_{ST}$ estimation, studies that simply rank the most highly differentiated SNPs between populations are unable to evaluate genome-wide significance of reported signals. On the other hand, model-based approaches (Bhatia et al. 2011; Ayodo et al. 2007; Lewontin and Krakauer 1973; Price et al. 2009a) can formally assess genome-wide significance and are robust to the biases of the WC $F_{ST}$ estimator at single SNPs. In general, studies that use a model-based approach have sufficient statistical power if sample sizes are much larger than $1/F_{ST}$ (Bhatia et al. 2011), as both $F_{ST}$ and sampling error contribute to normal variation in allele frequency differences. In the Jin et al. comparison, the sample size of Yoruba (n=113), is much smaller than the reciprocal of $F_{ST}$ between AAF and YRI ($1/F_{ST}$ = 1429). Thus, sampling variation is expected to dominate the variance of the distribution of allele frequency differences. This may contribute to the fact that none of the reported SNPs achieves genome-wide significance ($P < 5 \times 10^{-8}$) when re-evaluated using a model-based approach (Ayodo et al. 2007; Bhatia et al. 2011) (see Table 2).

We re-examined the statistical significance of the 10 novel loci reported by Jin et al. in the Bhatia et al. (2011) dataset, which included 6,209 African Americans and 756 Yoruba. (Extending the analysis to all 29,141 African Americans in the current study yields very similar results, as the Yoruba sample size is the limiting factor.) The Bhatia et al. data include 9 of these 10 loci, and only 4 of the 9 loci were nominally significant (P<0.05 without correcting for multiple hypothesis testing) (see Table 2). We caution however that these 4 loci should not be viewed as an independent replication, because the two analyses are not statistically independent due to genetic drift between AAF and YRI populations that is common to both analyses, so that loci in the tail of one analysis could be expected to lie in the tail of the other analysis. The lack of nominal significance at most loci in the non-independent analysis of Bhatia et al. data suggests that most of the reported novel loci are false positives, although a subset may be genuine.

It is important to recognize that even robust, genome-wide significant evidence of unusual population differentiation does not imply selection following the forced migration from Africa. For example, , at the 4 loci identified by both Bhatia et al. (2011) and Jin et al. (2012), the observed population differences are more parsimoniously explained by selection within Africa. This selection could have occurred in the ancestors of Yoruba and/or in the African ancestors of African Americans (prior to enslavement). This is because the majority of time since these populations diverged was spent in Africa, giving more time for selection to produce an allele frequency difference.

As a counterexample that proves the rule, we consider the well-studied sickle-cell variant rs334 at the HBB locus, where biological evidence suggests some selection since the arrival of Africans in the Americas is likely to have occurred. Homozygotes for the recessive allele are afflicted with sickle-cell anemia, a debilitating condition that results in very low fertility. However, the minor allele at rs334 is maintained at high frequency in Africa because heterozygotes have increased malaria resistance (Aidoo et al. 2002). The minor allele frequency at rs334 in African Americans is 0.050 (Auer et al. 2012), corresponding to an allele frequency of 0.063 (0.050/0.8) on African segments. The strongest possible negative selection against the minor allele would occur if have no advantage (due to much lower rates of malaria in the Americas) and that no people with sickle-cell anemia have children). Conservatively assuming this model, the allele frequency in the African ancestors of African Americans 7 generations ago (Price et al. 2009b) would have been 0.096 (see Methods).  This corresponds to a maximum possible allele frequency difference of 0.033 due to selection since the migration from Africa. Allele frequency differences at HBB of >10% have been reported between African populations (Ayodo et al. 2007; Bhatia et al. 2011), showing that selection in Africa cannot be ruled out as explaining most of the observed frequency difference between African Americans and Yoruba.  We have no doubt that the frequency of the minor allele decreased in African Americans since arrival in the Americas. However, the empirical data do not allow us to conclude that the allele frequency difference between African Americans and Yoruba at the sickle cell variant is primarily explained by selection since the arrival of Africans in the Americas.

We note in passing that the lack of a genome-wide significant deviation in average local ancestry at the HBB locus ($\gamma$ =0.206, vs. a genome-wide average of 0.204) is not unexpected, even given the decrease in frequency of this allele that must surely have occurred. Even though the per-allele selection coefficient is strong ($s_{allele} = 0.077$), the selection coefficient per copy of local ancestry is still quite low ($s_{ancestry} = 0.0074$), and below the threshold of $s_{ancestry}$ = 0.019 that our local ancestry analysis is powered to detect (see Methods). According to our model of selection we expect an average local ancestry of 0.210 at the HBB locus (see Methods), consistent with our observed result of $\gamma$ =0.206 (1.11 s.d. apart).

## Discussion

We performed a scan for unusual deviations in local ancestry in African Americans compared with the genome-wide average, which found no genome-wide significant loci and did not replicate 6 previously reported loci with unusual deviations in local ancestry. We also reanalyzed 14 unusually differentiated loci from a previous study using a different $F_{ST}$ estimator, showing that many single-SNP $F_{ST}$ estimates were inflated. Even after correcting for this inflation, most of the reported loci are not genome-wide significant. Furthermore, even for loci with robust, genome-wide significant evidence of selection based on population differentiation, selection within Africa provides a parsimonious explanation for most of the empirically observed differentiation.

We caution that although there is little evidence of selection since the forced migration out of Africa in the data analyzed by Jin et al. (2012) or in the current study, we cannot exclude the possibility that some selection of this type has occurred.  Indeed, selection is an ongoing process, and has surely occurred to some degree in African Americans since migration out of Africa. The key point here is statistical power. Our genome-wide scan of 29,141 samples study is well-powered (>95%) to detect signals of selection with a selection coefficient for local ancestry ( $s_{ancestry}$) greater than 0.019 (see Methods). Although selection of this magnitude or greater is ruled out by our data, weaker selection may have occurred. For example, selection after the forced migration from Africa is likely to have occurred at the HBB locus due to much lower rates of malaria and a corresponding reduction in positive selection. However, our estimate of the likely selection coefficient for local ancestry ($s_{ancestry}$ = 0.0074) is too small for us to produce genome-wide significant evidence of selection even in the context of the large sample size we analyzed.

We conclude with three recommendations for future studies. First, studies of selection since admixture based on deviations in local ancestry in African Americans or in other admixed populations with similar ages of admixture should employ a genome-wide significance threshold of $P=1 \times 10^{-5}$ (Seldin et al. 2011), and should be cognizant of the possibility that systematic errors in local ancestry inference can lead to false-positive signals. Second, studies of selection based on population differentiation that involve unequal sample sizes should not use the $F_{ST}$ estimator of Weir and Cockerham (1984), which is susceptible to bias in this case, and instead should use the Hudson estimator (Hudson et al. 1992; Weir and Hill 2002; Bhatia et al. 2013). Third, genome-wide significance should never be reported based on a simple ranking, and instead should be reported via model-based approaches (Lewontin and Krakauer 1973; Ayodo et al. 2007; Price et al. 2009a; Bhatia et al. 2011; Grossman et al. 2010).

## Methods

*Samples*

We studied the local ancestry distribution of African Americans from five different cohorts (N = 29,141 samples combined across cohorts), using a previously published data set where a nearly identical local ancestry inference procedure was used to study the rate of recombination (Hinch et al. 2011).

In detail, the dataset was derived from five genome-wide association studies conducted on African Americans, all of which differ in population size and characteristics. A coherent summary of the generation of these five datasets and the filters we used to harmonize the genotyping data is presented in (Hinch et al. 2011), and hence we do not present it again here.

A complication in the analysis of these data is that the data were produced on three different genotyping platforms. Three of these data sets are genotyped on the Illumina 1M array: samples from the African American Lung Cancer Consortium (AALCC), the African American Breast Cancer Consortium (AABCC), and t the African American Prostate Cancer Consortium (AAPCC). The fourth data set was genotyped on the Illumina Human Hap550 array and is from the Children's Hospital of Philadelphia (CHOP) (Hinch et al. 2011). The fifth data set is the Candidate Gene Association Resource (CARe) study, a consortium of cohorts. This data set consists of the ARIC, CFS, CARDIA, JHS and MESA cohorts and is genotyped on the Affymetrix 6.0 chip. We note that the AALCC, AABCC and AAPCC data sets consist of disease cases and controls, but phenotype information was not available in the current study. The inclusion of disease cases could produce false-positive signals of selection due to admixture associations to disease (no such signals were observed), but are unlikely to produce false-negative signals of selection.

For our local ancestry inferences of the CARe dataset, we simply used the already published results of Pasaniuc et al. (2011). All of the remaining datasets were curated to retain only markers and samples with high genotyping completeness (>90%). We removed samples with genetic evidence of being second-degree relatives or closer using either PLINK (Purcell et al. 2007) or EIGENSOFT (Patterson et al. 2006) (*smartrel)*. Samples with genome-wide European ancestry proportion less than 2.5% or greater than 75% were removed in all cohorts. All these datasets were separately combined with the phased Hapmap3 data of 88 European (CEU) and 100 West African (YRI) samples. Markers were removed if their allele frequency was inconsistent with a linear combination of 0.8 African and 0.2 European allele frequencies according to a t-statistic greater than 15 (or less than -15). This filter was applied to each cohort individually. A total of 626 markers were removed, none of which were located inside the 6 regions that were previously reported as being loci affected by natural selection since admixture (Jin et al. 2012). The markers that were removed had values of $(p_{AA}-p_{EUR})/(p_{AFR}-p_{EUR})$ that were either greater than 0.25 or less than 2.86. These extreme deviations from the expected admixture proportion of African and European ancestral allele frequencies are likely due to genotyping artifacts.

Following QC, the remaining samples were 4,094 AALCC samples genotyped at 877,881autosomal markers; 5,131 AABCC samples genotyped at 866,269 autosomal markers; 6,339 AAPCC samples genotyped at 867,658 autosomal markers; 7,368 CHOP samples genotyped at 480,029 autosomal markers; and 6,209 CARE samples genotyped at 738,831 autosomal markers. These five datasets have 121,511 autosomal markers in common, which we used to generate Figure S1.

*Inferring Local Ancestry*

For the CARe cohort, we used the exact same local ancestry inference reported in Pasaniuc et al. (2011). For the remaining datasets, we ran the HAPMIX (Price et al. 2009b) software separately in each cohort to infer local ancestry estimates for all the samples at each of the autosomal loci. The software was run using the Hapmap3 CEU and YRI haplotypes as the ancestral populations, assuming that the number of generations since admixture ($\lambda$) was 6, using an individual specific average European ancestry proportion ($\theta$) prior, and using the Oxford recombination map (Myers et al. 2005). The individual-specific $\theta$ values were calculated by running HAPMIX on these same samples using a uniform recombination map. The software was run in a mode in which it outputs an integer estimate of local ancestry by sampling from the probabilities for 0, 1 or 2 European chromosomes at each locus. The genome wide ancestry for each sample was calculated by averaging over local ancestry estimates genome-wide, after scaling these estimates by half. The average local ancestry at each locus was calculated as an average of the local ancestry estimates across all the samples. The entire analysis was conducted separately for each cohort and then combined across cohorts. Because of issues with ancestry inference at the ends of chromosomes, we removed the first and last 2 Mb of each chromosome from analysis. We note that 3 loci in these regions (which do not overlap with the Jin et al. loci) did show significant deviations in local ancestry, but these are very likely to be artifacts (see Supplementary Note). This filtering left a total of 118,006 SNPs in our analysis of local ancestry, which we used to generate Figure 1.

The mean European genome-wide ancestry proportion was 0.210 (S.D across individuals 0.123) in the AALCC sample, 0.218 (0.134) in the AABCC sample, 0.215 (0.131) in the AAPCC sample, and 0.193 (0.086) in the CHOP sample. These estimates are similar to previous studies (Pasaniuc et al. 2011; Smith et al. 2004). The standard deviation in average local ancestry estimates across SNPs was 0.0036 for the full set of 29,141 samples.

*Theoretical Standard Deviation in Local Ancestry*

We calculated the theoretical standard deviation in average local ancestry as follows:

$$SD^*(\gamma) = \frac{\sqrt{\sum_i 2\bar{\gamma}_i(1-\bar{\gamma}_i)}}{2N}$$

Where $\bar{\gamma}_i$ is the average genome-wide ancestry for individual *i*, and *N* is the total number of samples. Using this calculation we obtain a theoretical standard deviation of 0.0016, less than half the empirical standard deviation of 0.0036.

*Assessing Population Differentiation with the WC Estimator*

Consider a biallelic SNP in a sample of individuals from 2 populations. The WC estimator is:

$$\hat{F}_{ST}^{WC} = 1 - \frac{2\frac{n_1 n_2}{n_1+n_2}\frac{1}{n_1+n_2-2}[n_1\tilde{p}_1(1-\tilde{p}_1)+n_2\tilde{p}_2(1-\tilde{p}_2)]}{\frac{n_1 n_2}{n_1+n_2}(\tilde{p}_1-\tilde{p}_2)^2 + (2\frac{n_1 n_2}{n_1+n_2}-1)\frac{1}{n_1+n_2-2}[n_1\tilde{p}_1(1-\tilde{p}_1)+n_2\tilde{p}_2(1-\tilde{p}_2)]} \quad (1)$$

where $n_i$ is the sample size and $\tilde{p}_i$ is the sample allele frequency in population *i* for $i \in \{1,2\}$. Then, in the limit of large sample sizes ($n_i - 1 \approx n_i$), we can assume that sample allele frequencies become close to population allele frequencies ($\tilde{p}_i \rightarrow p_i$). We analyze the estimator as the sample sizes increase, but their ratio goes to a constant *M*. In the limit of infinite, but not necessarily equal sample sizes the estimator is:

$$\lim_{\substack{n_1,n_2 \rightarrow \infty \\ n_1/n_2 \rightarrow M}} \hat{F}_{ST}^{WC} = \frac{(p_1-p_2)^2}{(p_1-p_2)^2 + 2\frac{1}{(M+1)}[Mp_1(1-p_1)+p_2(1-p_2)]} \quad (2)$$

Consider a SNP that is rare in one population and has allele frequency zero in the other population: $p_1 = 0, p_2 = \varepsilon$. If sample sizes are equal, the single SNP estimate of $F_{ST}$ from the WC estimator is approximately $\varepsilon$. Now, consider what happens as we increase $n_1$ arbitrarily. It is clear that both numerator and denominator tend toward the same quantity and $F_{ST}$ approaches 1.

*Changes in Estimator Alter the 99.99th Percentile*

Use of the Hudson $F_{ST}$ estimator instead of the WC estimator results in lower estimates of $F_{ST}$ at the loci reported by Jin et al. However, it is possible that the 99.99th percentile threshold is also lowered by use of this estimator and that reported loci still fall at this upper tail of the distribution. To assess this effect in sample sizes similar to those of Jin et al. (2012) we sub-sampled 2,500 African

American individuals from our data, subtracted European allele frequencies from CEU (Bhatia et al. 2011), and compared the result to YRI using both the WC and Hudson $F_{ST}$ at every SNP. According to this analysis, the 99.99th percentile of $F_{ST}$ was 0.048 for the WC estimator and 0.046 for the Hudson estimator.

Jin et al. report a threshold of 0.0452. Even if this decreases by 0.002 due to use of the Hudson estimator, 2 of the 14 reported loci would no longer be in the 99.99th percentile.

*Selection at HBB after the migration of African American ancestors from Africa*

In order to assess the maximum effect of selection at HBB, we consider the following situation. The minor allele at rs334—which is known to cause sickle cell anemia—in African Americans today has a frequency of 0.050 (Auer et al. 2012), corresponding to a MAF of 0.0625 (0.05/0.8) on African segments. From this information, we can work backwards in time with the following equation:

$$p_{g+1} = \frac{p_g}{1-p_g} \qquad (3)$$

Assuming that $p_0 = 0.0625$, and 7 generations since the admixture of the African and European ancestors of African Americans (Price et al. 2009b), we have $p_7 = 0.0962$. According to these estimates, the maximum allele frequency difference since admixture is 0.0337.

Under this model, the per-allele selection coefficient is simply the allele frequency in the population—not on African segments alone—at the current generation ($s^g_{allele} = \gamma p_g$), where $\gamma$ is the proportion of African ancestry at the HBB locus during the current generation. Assuming that the proportion of local ancestry at each locus 7 generations ago is equivalent to the current genome-wide average, the maximum value of this coefficient is $s_{allele} = 0.796 p_7 = 0.077$. The selection coefficient per copy of African local ancestry is given by $s_{ancestry} = \gamma(p)^2$. That is, given that an individual carries one African chromosome at the HBB locus he must also carry (1) the sickle allele on this first African chromosome (with probability *p*) (2) a second African chromosome at this locus (with probability $\gamma$) and (3) the sickle cell allele on that second African chromosome (with probability *p*). According to our model, the maximum value of this coefficient is $s_{ancestry} = 0.796(p_7)^2 = 0.0074$.

*Estimating Local Ancestry Proportions After Selection*

To perform power calculations and estimate the expected deviation in local ancestry at HBB (see above) we need to be able to assess the effect of a particular selection coefficient on local ancestry proportion. We did this iteratively, using the equation

$$\gamma_{g+1} = \frac{\gamma_g(1-s_{ancestry})}{1-\gamma_g s_{ancestry}} \quad (4)$$

We performed this iteration $g$ times to assess the effect of selection at the locus. To perform power calculations, we used a static value of $s_{ancestry}$, to assess the expected deviation in local ancestry at HBB we substituted $s_{ancestry} = \gamma_g(p_g)^2$.

*Estimating the Minimum Detectable Selection Coefficient*

In order to call a genome-wide significant deviation in local ancestry, we must have $|\hat{\gamma}_L - \bar{\gamma}| > 4.4 \hat{\sigma}_{\gamma_L}$. That is, the observed average ancestry at a locus $\hat{\gamma}_L$, must deviate from the genome wide average $\bar{\gamma}$ by more than 4.42 standard deviations (estimated from the data). We can assume that the sampling distribution of observed average local ancestry is normal and centered around the true population average ancestry at the locus. That is $\hat{\gamma}_L \sim N(\gamma_L, \frac{\gamma_L(1-\gamma_L)}{n})$ where $\gamma_L$ is the true population average ancestry at the locus and $n$ is sample size of the study. We can then solve for $\gamma_L$, so that $\Pr(|\hat{\gamma}_L - \bar{\gamma}| > 4.4 \hat{\sigma}_{\gamma_L} | \gamma_L) = 0.95$. In our case, assuming $\bar{\gamma} = 0.204$, and $\hat{\sigma}_{\gamma_L} = 0.0036$ we obtain $\gamma_L$ = 0.183 or 0.225. Then, assuming 7 generations since admixture we perform a grid search over possible values of the selection coefficient for local ancestry that would produce these values of $\gamma_L$ (see below) and obtain an estimate of 0.019.

*Using a model-based approach to detect selection on Jin et al. (2012) data*

When using a model-based approach to reanalyze Jin et al. results, we were unable to estimate $F_{ST}$ in their sample (since we did not have the raw genotypes) and thus used their reported $F_{ST}$ of 0.0007 in the model (Ayodo et al. 2007; Price et al. 2009a; Bhatia et al. 2011; Lewontin and Krakauer 1973). We note that this $F_{ST}$ was calculated using the WC estimator, which may be susceptible to biases when very different sample sizes are analyzed. The reported $F_{ST}$ of 0.0007 was less than the $F_{ST}$ of 0.0011 used in Bhatia et al. (2011). This difference has a minimal impact on the resulting statistics, as the variance is primarily due to the small sample size from Yoruba. However, using an $F_{ST}$ of 0.0011 would lead to even less statistically significant results than those reported in Table 2, so that all model-based P-values using Jin et al. data would remain non-genome-wide significant.

## Acknowledgements

We thank N. Patterson, and J. Wilson for helpful discussions and comments on the manuscript. This research was funded by NIH grants R01 HG006399 and R03 HG006170.

**Table 1.** We list the 6 regions with unusual deviations in local ancestry reported by Jin et al. and compare these to our scan. None of the 6 regions replicated at nominal significance (P < 0.05) in our analysis.

| Region | Jin et al. | | Current study | |
| --- | --- | --- | --- | --- |
| | Deviation | Nominal P-Value | Deviation | Nominal P-Value |
| chr1:17409539..21604321 | -0.025 | 7.43E-04 | -0.004 | 0.55 |
| chr2:241750403..242568618 | -0.023 | 2.07E-03 | -0.006 | 0.44 |
| chr2:37451925..37508581 | 0.023 | 2.16E-03 | 0.005 | 0.51 |
| chr3:116930811..118313302 | 0.025 | 8.58E-04 | -0.002 | 0.83 |
| chr6:163653158..163653428 | 0.023 | 2.70E-03 | 0.004 | 0.60 |
| chr16:61214438..61242497 | 0.023 | 2.26E-03 | 0.006 | 0.41 |

**Table 2.** We recreate Table 2 of Jin et. al (2012) analyzing the same data with the Hudson instead of the WC estimator. The bolded cells indicate loci that fall below the 99.99th percentile threshold of 0.0452 when the Hudson estimator is used. We also estimated the P-value at each SNP using the reported $F_{ST}$ = 0.0007 of Jin et al. (2012) (see Methods), and a model based approach (Ayodo et al. 2007). Finally, we report the model-based P-value of the most significant SNP in the region reported in the parallel study of Bhatia et al. (2011). We note that results reported in that paper were more significant than those reported here due to analysis of additional populations. The chr16 locus is reported as N/A due to a lack of data at this locus in the Bhatia et al. data.

| SNP id | Region | Gene | WC $F_{ST}$ Jin Data | Hudson $F_{ST}$ Jin Data | Model-based P-value Jin Data | Model-based P-value Bhatia Data |
|---|---|---|---|---|---|---|
| rs1541044 | chr1:100125058..100183875 | | 0.0562 | **0.0439** | 4.7 x $10^{-5}$ | 0.04 |
| rs4460629 | chr1:153401959..153464086 | | 0.0692 | 0.065 | 6.8 x $10^{-7}$ | 2.1 x $10^{-4}$ |
| rs12094201 | chr1:236509336 | | 0.0561 | 0.0489 | 1.7 x $10^{-5}$ | 0.86 |
| rs7642575 | chr3: 31400165 | | 0.0453 | **0.0393** | 1.1 x $10^{-4}$ | 0.41 |
| rs652888 | chr6:26554684..33961049 | HLA | 0.0711 | 0.0627 | 1.1 x $10^{-6}$ | 1.8 x $10^{-11}$ |
| rs9478984 | chr6:151555551..151569258 | | 0.0545 | 0.0596 | 2.1 x $10^{-6}$ | 0.02 |
| rs10499542 | chr7: 22235870 | | 0.0461 | 0.0453 | 3.6 x $10^{-5}$ | 0.35 |
| rs304735 | chr7:79768487..80482597 | CD36 | 0.0946 | 0.069 | 3.0 x $10^{-7}$ | 3.7 x $10^{-13}$ |
| rs2920283 | chr8:143754039..143758933 | PSCA | 0.0468 | 0.0532 | 7.6 x $10^{-6}$ | 6.4 x $10^{-7}$ |
| rs1498487 | chr11:5034229..5421456 | HBB | 0.0617 | 0.0464 | 2.4 x $10^{-5}$ | 1.7 x $10^{-7}$ |
| rs4883422 | chr12:7189594 | | 0.0472 | 0.0461 | 3.0 x $10^{-5}$ | 1.3 x $10^{-3}$ |
| rs6491096 | chr13:25488362 | | 0.0472 | **0.0373** | 1.5 x $10^{-4}$ | 0.4 |
| rs1075875 | chr16: 47595721 | | 0.0766 | 0.0608 | 1.3 x $10^{-6}$ | N/A |
| rs6015945 | chr20:59319574 | | 0.0627 | 0.055 | 4.3 x $10^{-6}$ | 0.5 |

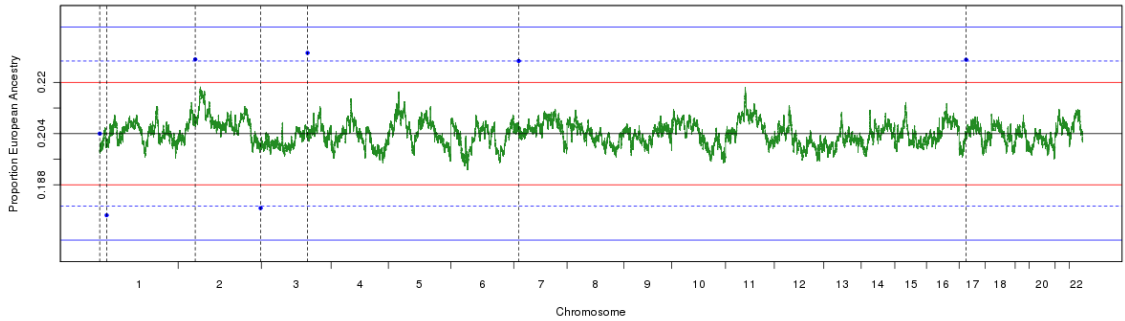

**Figure 1. Ancestry at each location in the genome in 29,141 African Americans.** This figure gives the proportion of European ancestry at each of the 118,006 SNPs common to all cohorts. The black line indicates the genome-wide average proportion European ancestry. The red and blue lines indicate the threshold for genome-wide significance ($P < 10^{-5}$) in our study, and the Jin et al. study, respectively. The dashed blue line indicates the threshold for significance ($P < 2.7 \times 10^{-3}$) that was actually used in the Jin et al. study. The standard deviation was computed empirically over all SNPs. It is clear that no region attains genome-wide significance in our scan. Dashed vertical lines indicate the location and blue points the deviation in local ancestry of the six loci reported under selection in Jin et al. These deviations are reported relative to the genome-wide average ancestry proportion in our study. None of the six reported loci exceed the $P < 10^{-5}$ genome-wide significance threshold for the Jin et al. study (blue lines).